\def\etal{{\rm et al. }}

\def\kpc{{\ \rm kpc}}
\def\kms{{\ \rm km\, s^{-1}}}

\documentclass[usenatbib]{mn2e}

\usepackage{amssymb}
\usepackage{graphicx}
\usepackage{bmpsize}
\usepackage{cite}
\usepackage{natbib}
\usepackage{hyperref} 
\usepackage{aas_macros}
\usepackage{times}
\usepackage{epsfig}
\usepackage{amsmath}
\usepackage{natbib}
\usepackage{longtable}
\usepackage{psfrag}
\usepackage{txfonts}
\usepackage{pdflscape}
\usepackage{subfigure}
\usepackage{xcolor} 
\begin{document}

\title[AGNs in Small Galaxy Systems]{AGNs in Small Galaxy Systems: comparing the main properties of active objects in pairs, triplets and groups.}

\author[Duplancic \etal]{Fernanda Duplancic$^{1}$\thanks{E-mail: fduplancic@unsj-cuim.edu.ar}, Diego G. Lambas$^{2}$, Sol Alonso$^{1}$  \& Georgina V. Coldwell$^{1}$ \\ 
$^{1}$ Departamento de Geof\'{i}sica y Astronom\'{i}a, CONICET, Facultad de Ciencias Exactas, F\'{i}sicas y Naturales, Universidad Nacional \\
de San Juan, Av. Ignacio de la Roza 590 (O), J5402DCS, Rivadavia, San Juan, Argentina\\
$^{2}$ Instituto de Astronom\'ia Te\'orica y Experimental (IATE-CONICET), Laprida 854, C\'ordoba, Argentina }

\date{\today}

\pagerange{\pageref{firstpage}--\pageref{lastpage}}

\maketitle
\label{firstpage}

\begin{abstract}

We perform a comparative study of AGNs in pairs, triplets and groups. To this end we use the Duplancic et al. catalogue of small galaxy systems and consider BPT and WHAN diagnostic diagrams to select optical AGNs. Also we identify mid-IR AGNs by using WISE data. We performed a comparison between the different AGN classification methodologies and study the AGN fraction in pairs, triplets, and groups with four to six members. We also analyse the main properties of Optical and mid-IR AGN hosts and the influence of environment on the active nuclei phenomena in these small galaxy systems. Our results show that, regardless the specifically adopted classification scheme, the fraction of AGN in pairs and triplets is always higher than the corresponding fraction in groups. Moreover, the fraction of powerful AGNs in pair and triplets is about twice the fraction of regular AGNs. We also find a remarkable difference between Optical and mid-IR AGNs in groups, where host  galaxies of WISE AGNs are less massive and concentrated, with young stellar populations and blue colours. Also all WISE AGNs in groups have a very close companions and reside in an intermediate global density environment. Galaxy triplets show a larger AGN fraction for galaxies with a close nearest neighbours, while pairs present a nearly constant AGN fraction regardless the distance to the nearest companion. Our studies highlight the important role of interactions, besides the  global environment dependence, in the activation of the AGN phenomenon in small galaxy systems.
 
\end{abstract}
\maketitle

\section{Introduction}
\label{Intro}
Interactions between galaxies are an efficient mechanism to feed the central black hole in active nuclei galaxies. When two galaxies interact, gravitational instabilities may generate gas flow to the inner most central regions, providing the fuel for the growth by accretion of central black holes \citep[e.g,][]{Sanders1988,Storchi-Bergmann2001,Alonso2007,Ellison2011,Satyapal2014,HernandezIbarra2016,Storchi-Bergmann2019}. Several works have shown that the most luminous active galaxies are preferentially hosted by major mergers \citep[e.g][]{Urrutia2008,Treister2012,Glikman2015}, in agreement with theoretical studies. In lower luminosities, minor mergers feed central black holes in early galaxies, while secular processes dominate in gas-rich galaxies \citep{Simoes-Lopes2007,Neistein2012}. In addition,  based on observational evidence, different studies have found a significant increase of the nuclear activity in less-luminous AGN galaxies with tidal interaction features or distorted morphologies with respect to AGN hosts without signs of interactions \citep[e.g.][]{Koss2010,Koss2012,Ellison2011,Ellison2013,Silverman2011,Sabater2013}. In this direction, \citet{Alonso2007} carried out a statistical analysis comparing AGN galaxies in close pairs with active objects in an isolated environment. The authors found that for galaxies with strong interaction features, the nuclear activity and accretion rate of the AGN are significantly larger than for active galaxies without near companions. \citet{Sabater2015} suggested an indirect way in the effect of AGN activity by galaxy interactions, influencing the central gas supply.  More recently,  \citet{Barrows2017}  found that the increases in star formation activity are correlated with enhanced AGN luminosity, suggesting that both values are mutually triggered by mergers and interactions. All these studies provide us obvious clues about AGN fuelling and its link with galactic interactions.

\citet{Coldwell2014} analyse different properties of the small-scale environment of optically selected Seyfert 2 finding that active galactic nucleus occurrence is higher in lower/medium density environments with a higher merger rate and a lack of a dense intergalactic medium that can strip gas from these systems. This scenario may provide suitable conditions for the central black hole feeding.
\citet{Koulouridis2006} presented a study of the local and the large scale environment of bright IRAS galaxies, with the aim to study the active galactic nucleus-starburst connection. The authors found that close interactions can drive molecular clouds toward the central region of the galaxies, triggering starburst and obscuring the nuclear activity. When the galaxy neighbour moves away, an obscured type 2 AGN appears and the starburst is reduced. The decoupling of the galaxy pair gives birth to an unobscured (type 1) active galactic nucleus. Moreover, \citet{Koulouridis2013} analyse optical spectroscopy and X-ray imaging of neighbouring galaxies around samples of Seyfert 1 and Seyfert 2. More than 70\% of the nearby galaxies exhibited star formation and/or nuclear activity, while X-ray analysis showed that this percentage may be higher, indicating a link between close interactions, and star formation/nuclear activity. 
More recently \citet{Coldwell2017} found that LINERs are more likely to populate low-density environments in spite of their morphology, which is typical of high-density regions such as rich galaxy clusters. In this context galaxy groups which are compact, i.e. with close members and a low velocity dispersion represent an ideal environment promoting galaxy-galaxy interactions and mergers that can trigger nuclear activity. 

The AGN content in compact groups have been study by different authors. \citet{Coziol1998} study 82  brightest  galaxies  in  a  sample  of  17 compact groups finding that 38\% of the objects have an AGN  and this percentage reaches 47\% at the core of the groups. They also found that 50\% of the AGN population in these groups are low-luminosity AGN galaxies. \citet{Martinez2010} study intermediate-resolution optical spectra of 270 member galaxies, in 64 compact groups finding a high fraction of 68\% galaxies with AGN activity. The authors also report that most AGN have a low luminosity and find no dusty AGN host galaxies in their galaxy sample of compact groups. For the Compact group SDSS J0959+1259 \citet{DeRosa2015} also found a high percentage of active galaxies, with 60\% of group members hosting an AGN. Nevertheless, \citet{Sohn2013} suggest that the AGN fraction of compact group galaxies depends on the AGN classification methods, finding percentages ranging from 17\% to 42\% in a sample of 238 galaxies in 58 compact groups. Based on a  multi wavelength study (UV–IR) of a large sample of galaxies (7417 galaxies in 1770 compact groups) \citet{Bitsakis2015} found that late-type galaxies in dynamically evolved compact groups have a 15\% increment in the AGN population. In comparison to isolated galaxies and interacting pairs, the authors found no differences in the properties of nuclear activity. 
All the studies described above suggest that the absence of powerful AGN in compact groups are consistent with a scenario of gas depletion  related to tidal stripping and a consequent detriment on the supermassive black hole accretion rate. 
For galaxy triplets \citet{Duplancic2013} found that these systems can be considered as an extension of compact groups with a lower number of members. Therefore as in the case of compact groups \citet{CostaDuarte2016} found a low fraction of strong AGN galaxies in triplets with most members classified as passive or retired, according to the WHAN diagnostic diagram.

Whether quenched or enhanced, the nuclear activity in a galaxy is strongly related to the molecular gas (i.e. the fuel). In small galaxy systems different processes may affect the formation, destruction, spatial redistribution and local density variations of this component. Therefore, understanding how gas flows into central regions of galaxies in these systems is key to a understand the evolutionary pathways of active galaxies in these environments. To this end in the current work we  present a study of nuclear activity in galaxies from the catalogue of \citet{Duplancic2018} (hereafter D18). This catalogue is based on a selection criteria that is homogeneous in the identification of systems with a low number of members (two to six) populating  environments that promote galaxy-galaxy interactions. Our aim is to perform a comparative study of the AGN population in pairs, triplets and groups. 

The paper is organised as follows: In section \ref{data} we describe the galaxy catalogues used in this work. A description of the different AGN selection criteria adopted in this work is detailed in section \ref{AGNselection}. In section \ref{results} we present the results of different studies, including the comparison between the different AGN classification methods and the study of the AGN fraction in pairs, triplets and groups. In this section we also analyse the main properties of Optical and mid-IR AGN hosts and the influence of environment on the active nuclei phenomena in small galaxy systems. A discussion of our findings in presented in section \ref{disc}. Finally in section \ref{conc} we summarise the main results of this work and present our conclusions.

Throughout this paper we adopt a cosmological model characterised by the parameters $\Omega_m=0.3$, $\Omega_{\Lambda}=0.7$ and $H_0=70~h~{\rm km~s^{-1}~Mpc^{-1}}$.

\section{Data}
\label{data}

The sample of galaxies were drawn from the Data Release 14 of Sloan Digital Sky Survey\footnote{https://www.sdss.org/dr14/} \citep[SDSS-DR14,][]{Abolfathi2018,Blanton2017}. This survey includes imaging in 5 broad bands ($ugriz$), reduced and calibrated using the final set of SDSS pipelines. The SDSS-DR14 provides spectroscopy of roughly two millions extragalactic objects including objects from 
the SDSS-I/II Legacy Survey \citep{Eisenstein2001,Strauss2002}, the Baryon Oscillation Spectroscopic Survey \citep[BOSS,][]{Dawson2013} and the extended-BOSS  \citep[eBOSS][]{Dawson2016}, from SDSS-III/IV.

In this work we consider the Legacy survey and obtain all data catalogues through $\rm SQL$ queries in the  publicly  available  Catalog  Archive  Server (CAS)\footnote{http://skyserver.sdss.org/casjobs/}. Photometric properties were taken from \texttt{Galaxy} view and spectroscopic information from \texttt{SpecObjAll} table. We have selected model magnitudes extinction corrected. These model magnitudes are k--corrected and appropriate for extended objects to provide reliable galaxy colours. k-corrections were derived using the empirical k-corrections presented by \citet{OMill2011} and we restrict our analysis to galaxies with $r$-band magnitudes in the range $13.5<r<17.77$. This range avoids saturated stars as well as assures spectroscopic completeness in SDSS Main Galaxy Sample \citep[MGS,][]{Strauss2002}. We apply the apparent magnitude cut after Galactic extinction correction, in order to obtain an uniform extinction-corrected sample.

 For the optical AGN selection, we use the publicly available SDSS emission-line fluxes taken from the MPA/JHU VAGC\footnote{Available in CAS \texttt{galSpecIndx, galSpecInfo, galSpecLine and galSpecExtra} tables}. The method for emission-line measurement is detailed in \citet{Tremonti2004} and \citet{Brinchmann2004}. 
To select AGN in the infrared we used the methodology described in \citet{Assef2018} based on the Wide-field Infrared Survey
Explorer \citep[WISE\footnote{The cross-matched WISE data is available in CAS \texttt{WISE\_allsky} and  \texttt{WISE\_xmatch} tables}; ][]{Wright2010}. This survey comprises mid-IR images of the entire sky in four bands, centred at 3.4, 4.6, 12, and 22$\mu m$, hereafter W1, W2, W3, and W4, respectively. The mid-IR AGN catalogues are based on AllWISE Data Release \citep{Cutri2013}.

\subsection{Small galaxy System Catalogue}
\label{catalogue}
In this work we use the catalogue of small galaxy systems presented in D18 constructed by using spectroscopic and photometric data from SDSS-DR14. For a detailed description of this catalogue we refer the reader to D18. 

Briefly, the selection criteria are homogeneous in the identification of systems with two to six members. Galaxies in these systems are within the redshift range $0.05\le \rm z\le 0.15$ and have absolute r-band magnitudes brighter than $\rm M_{\rm r}$ $=-19$. Also, galaxy members within the system are close in projection in the plane on the sky ($\rm r_{\rm p}\le \rm 200 \kpc$) and have radial velocity differences $\Delta{\rm V}\le 500 \kms$. For the identification of member galaxies in D18 systems we take into account the fibre collision effect in the SDSS spectroscopic data by including photometric information of close galaxies. Therefore interactions and merger events are expected in this sample. Moreover, it is worth to notice that 90\% of galaxies in D18 sample have spectroscopic information. The systems are compact (with a compacticity similar to Hickson compact groups) and the r-band absolute magnitude difference between the brightest and faintest system galaxy is less than 2 magnitudes, therefore galaxy members have similar luminosities.
Regarding the environment, D18 systems are locally isolated within a fixed circular aperture of 500~$\rm kpc$ radius and a radial velocity cut of 700 $\kms$.

 In this work we consider only spectroscopic galaxies in the D18 sample, comprising a total of 20641 galaxies out of which 18218 reside in pairs, 2090 in triplets and 333 in groups with 4 or more members. 
 
\section{AGN selection}
\label{AGNselection}

To identify AGN host galaxies (hereafter AGNs) in our sample, we use three methods. Two of these are based on spectroscopic optical data using emission-line ratios, the third method uses infrared photometry which is sensitive to optically obscured AGNs.

\subsection{BPT AGNs}
\label{BPTAGN}

\begin{figure}
  \centering
  \includegraphics[width=.45\textwidth]{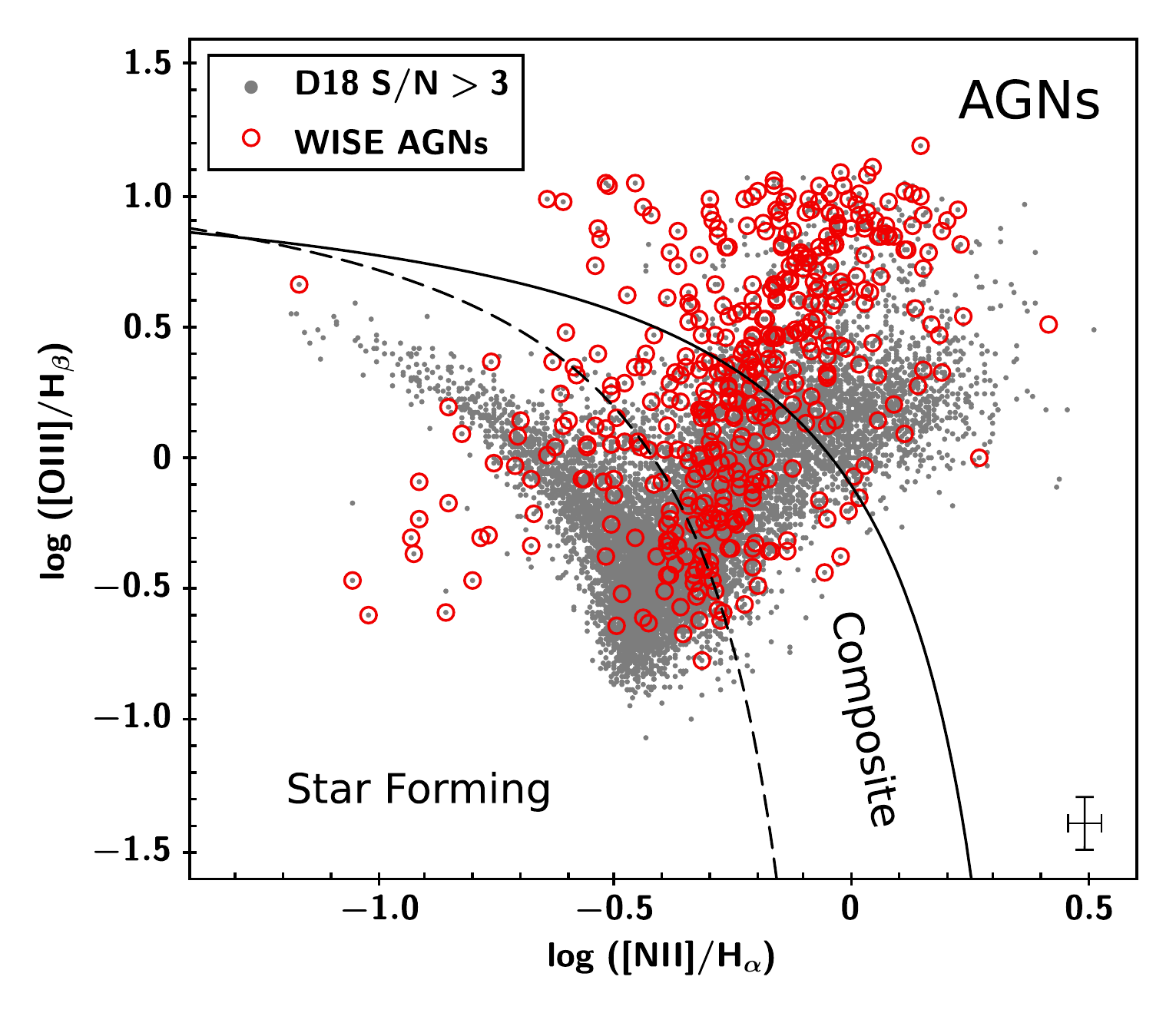}
  \caption{BPT diagram for D18 galaxies with S/N$>3$ in all four emission lines. Mean errors are represented at the bottom-right region of the plot. The solid line corresponds to the AGN demarcation from \citet{Kewley2001} and in dashed the \citet{Kauffmann2003} line for pure star-forming galaxies.  WISE AGNs are plotted as open circles.} 
\label{BPTdiagram}
\end{figure}

In order to identify  AGN  in the  sample of galaxies in small systems we use the  standard  diagnostic diagram proposed by \citet{BaldwinPhillipsTerlevich1981}, hereafter BPT. This diagram allows the separation of type 2 AGN from normal star-forming galaxies using emission-line ratios. Furthermore, we used only galaxies with signal to noise ratio S/N$>$3 for [OIII]$\lambda$5007 (hereafter [OIII]),  [NII]$\lambda$6583 (hereafter [NII]), H$\alpha$ and H$\beta$. From our catalogue of small galaxy systems, 10798 (52\%)  galaxies  fulfil this signal to noise restriction. Taking into account the relation between spectral lines, [OIII]/H$\beta$ versus [NII]/H$\alpha$, we find the BPT diagram shown in Fig.~\ref{BPTdiagram}. We classify galaxies in our sample according to their relative position in this diagram according to \citet{Kauffmann2003}. Therefore, we consider Composite galaxies in our AGN classification as objects residing between \citet{Kauffmann2003} and \citet{Kewley2001} demarcation lines. These Composite spectra  are expected to exhibit a mixture of star forming and AGN emission features \citep{Kewley2006}.

From this analysis we found 5105 AGNs, 4498 in pairs, 531 in triplets and 76 in groups. If we distinguish between pure AGNs and Composite objects we obtain 2136 pure AGNs (1854 in pairs, 243 in triplets and 39 in groups) and 2969 Composite galaxies (2644 in pairs, 288 in triplets and 37 in groups).

\subsection{WHAN AGNs}
\label{WHANAGN}

\begin{figure}
  \centering
  \includegraphics[width=.45\textwidth]{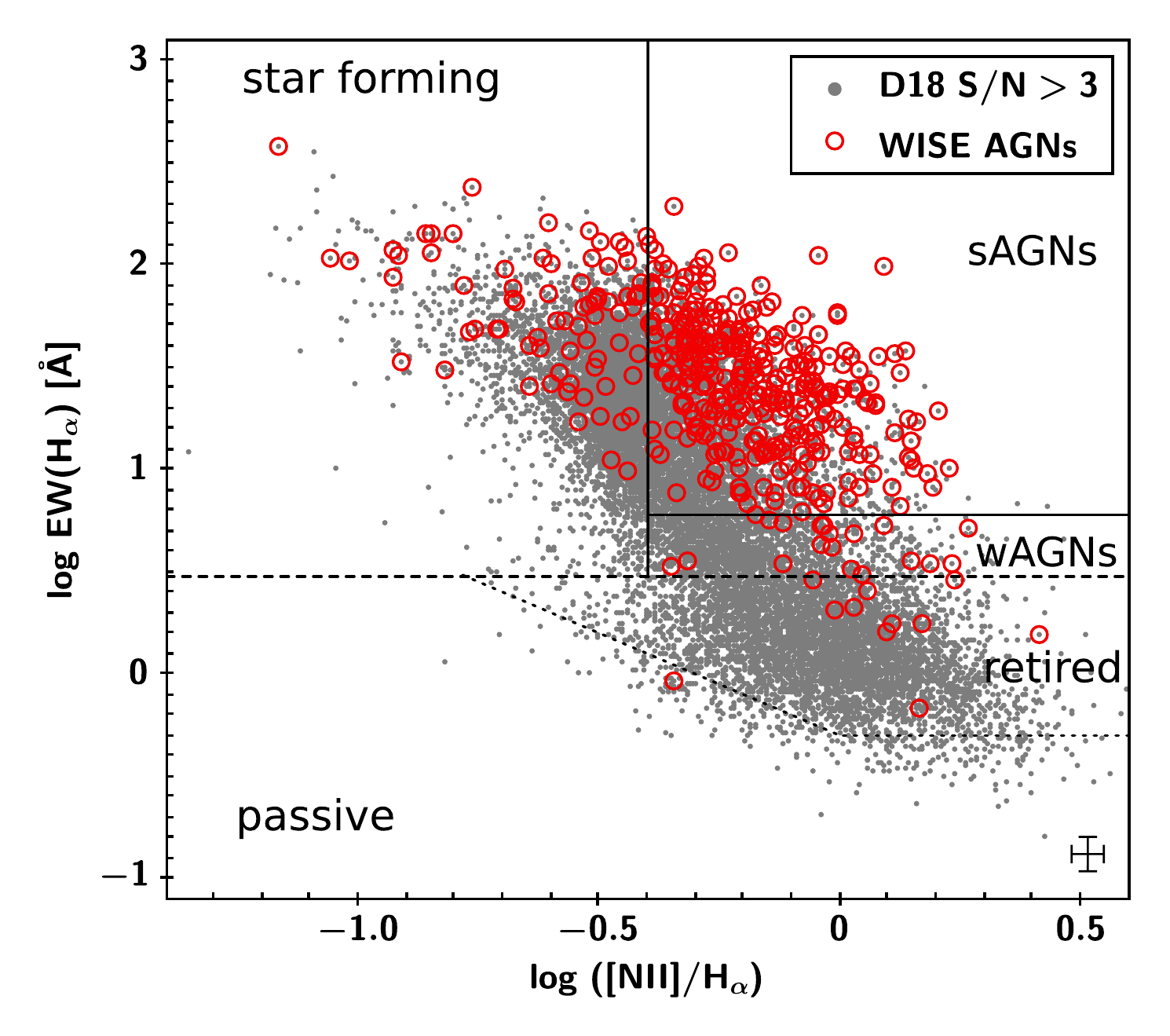}
   \caption{WHAN diagram for D18 galaxies with S/N$>$3 in [NII] and $\rm H_\alpha$. Mean errors are represented at the bottom-right region of the plot. We show demarcation lines from \citet{CidFernandes2011} for strong and weak AGNs as well as Star forming, Retired and passive galaxies.  WISE AGNs are plotted as open circles.} 
\label{WHANdiagram}
\end{figure}

Although the BPT diagram  is widely used for AGN classification, it is no able to disentangle
the so-called retired galaxy population from weak AGN types. Retired are galaxies without significant [OIII] emission, whose ionization mechanism  is usually explained by Hot Low-Mass Evolved Stars \citep[HOLMES;][]{Binette1994}. 
In this line  \citet{CidFernandes2011} showed that the WHAN diagram, i.e. the equivalent width  of H$\alpha$ (EW(H$\alpha$)) vs [NII]/H$\alpha$ can be used to separate the retired galaxy population from weak AGN types. Moreover, since only two lines are involved in this diagram, signal to noise restriction are lower and a higher percentage of emission line galaxies can be classified compared to the BPT approach. For the WHAN diagram we consider a signal to noise  S/N$>$3 for the used lines, finding that 15969 (77\%) galaxies in the sample of small systems fulfil this restriction. This percentage is 20\% higher than for the objects used in the BPT classification. In Fig.\ref{WHANdiagram} we show the WHAN diagram for our sample of galaxies in small systems and classify them into strong and weak AGN following \citet{CidFernandes2011}. Also, we identify retired and passive objects.

We find that 5685 galaxies are AGNs, 5061 are in pairs, 553 in triplets and 71 in groups. From this objects 3967 are strong AGNs (3559 in pairs, 367 in triplets and 41 in groups) and 1718 weak AGNs 
(1502 in pairs, 186 in triplets and 30 in groups).

\subsection{WISE AGNs}
\label{WISEAGN}

It is expected that a considerable fraction of the optical AGN emission is absorbed by the dust surrounding the central super massive black hole. This radiation is re-emitted at IR wavelengths and therefore, surveys at mid-infrared such as WISE are key to identify optically obscured AGN. There are several WISE colour diagnostics used to select AGN galaxies \citep{Mateos2012,Stern2012,Assef2013}. Most of them are based on the fact that  AGNs present a (W1-W2) colour redder than non-active galaxies because warm dust emission is more important than the light of the old stellar population in the host galaxy \citep{Stern2012,Assef2013}. Moreover, at low redshifts, the (W1-W2) colour index is less affected by extinction, making  AGN selection criteria based on (W1-W2) an efficient tool to select obscured AGN.

In this work we  apply  the  75\%  reliability  criteria presented in \citet{Assef2018} in order to select AGNs based on WISE data (hereafter WISE AGNs). Therefore, active objects must fulfil

 \begin{equation*}
(W1-W2) > \left\lbrace
\begin{array}{ll}
0.486 \textup{ exp}[0.092(W2-13.07)^2] & \textup{if } W2>13.07 \\
  0.486 &  \textup{if } W2\le13.07
\end{array}
\right.
\end{equation*}

 It is expected that more than 75\% of objects selected under these restrictions have bolometric luminosities dominated by the AGN \citep[for a detailed description see][]{Assef2018}.
  
\begin{figure}
  \centering
  \includegraphics[width=.45\textwidth]{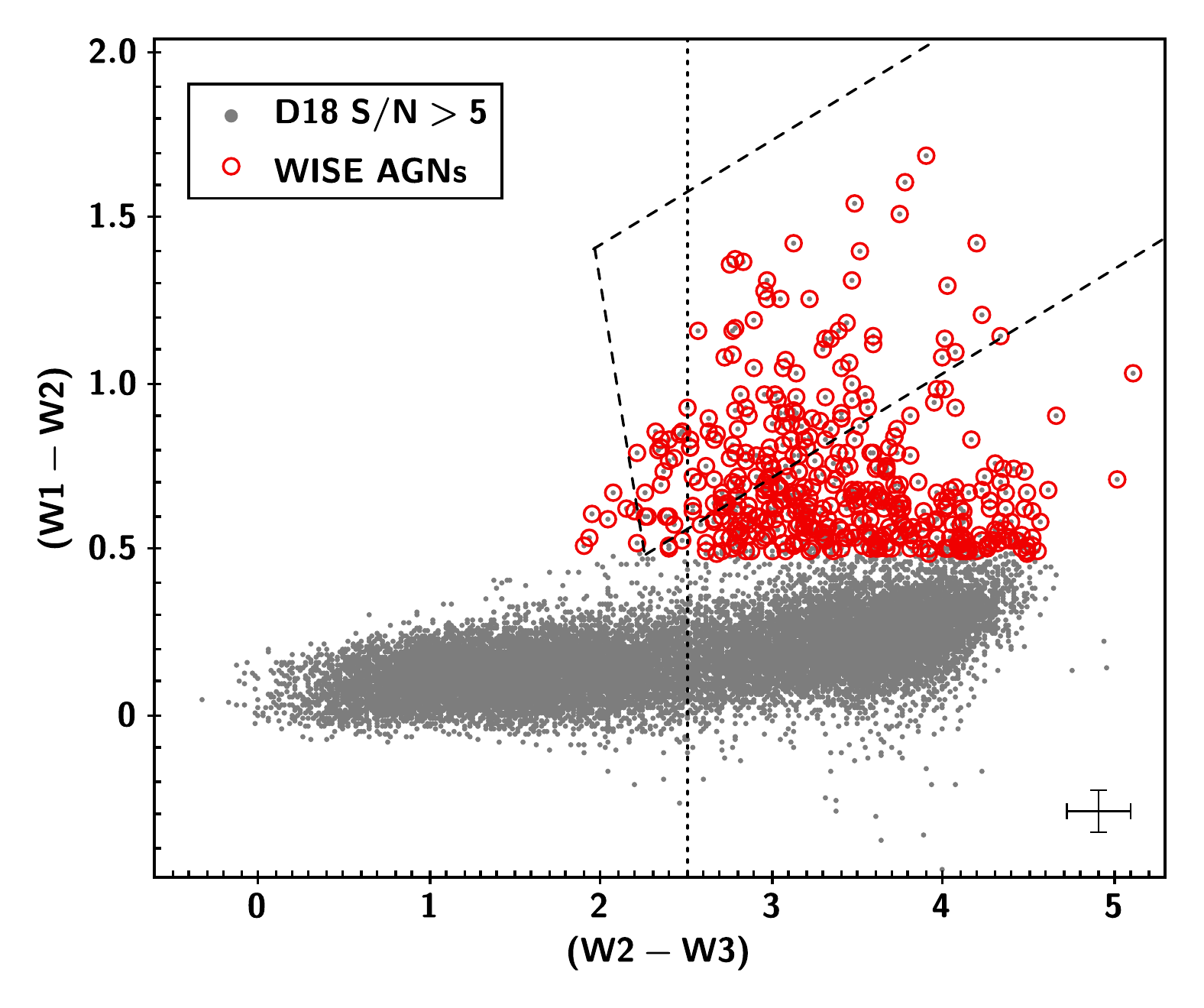}
  \caption{Mid-infrared colour–colour diagram of D18 WISE source with S/N$>5$ in W1 and W2. Mean errors are represented at the bottom-right region of the plot. WISE AGNs are plotted as open circles. The dashed lines illustrate the AGN selection wedge as defined by \citet{Mateos2012}. The vertical dotted line represents the limit suggested by \citet{Herpich2016} to separate between galaxies with current star formation from lineless galaxies.} 
\label{WISEcolorcolor}
\end{figure}

  By considering these criteria  we identify 434 AGNs (377 in pairs, 51 in triplets and 6 in groups).  It is important to highlight that 20379 (99\%) of the galaxies in the sample of small galaxy systems have W1 and W2 with signal to noise S/N$>$5. In Fig.\ref{WISEcolorcolor} we show W1-W2 versus W2-W3 colour-colour diagram for galaxies in our sample of small galaxy systems. It can be seen that mid-IR AGNs are located in a well defined region of this diagram. Also, for comparison we plot the AGN locus from \citet{Mateos2012}, this region corresponds to luminous AGNs based on X-ray data. We find that 134 (31\%) of 434 MIR AGNs identified in our sample are within this region. The vertical dotted line in Fig. \ref{WISEcolorcolor} represents the limit suggested by \citet{Herpich2016} to separate between galaxies with current star formation (W2-W3$>$2.5) from lineless galaxies (W2-W3$<$2.5).

\section{Results}
\label{results}

 In the following analysis we will compare the different methods used to identify AGNs in small galaxy systems. We will study the fraction of active objects in pairs, triplets and groups and its dependence on AGN power, host properties and environment.

 \begin{table*}
\centering
\caption{ Number of galaxies in pairs, triplets and groups and number of AGNs in these systems according to BPT, WHAN or WISE AGN classification. The values [*] correspond to the final BPT AGN classification without Retired/passive objects. We also include the number of Optical AGNs and the number of active objects in small galaxy systems classified  according to the combined criteria described in section \ref{compare}.} 
\begin{tabular}{lcccccc}
\hline\hline\noalign{\smallskip}
Name & galaxies & BPT[*]  & WHAN & Optical & WISE  &  Combined AGNs \\
\hline\noalign{\smallskip}
Pairs &  18218 & 4498(25\%)[3017(17\%)] & 5061(28\%) & 4814(26\%) & 377(2.0\%) & 5191(28\%)\\
Triplets & 2090 & 531(25\%)[328(16\%)]  &  553(26\%) & 520(25\%) &  51(2.4\%) &  571(27\%)\\
Groups  & 333 & 76(23\%)[43(13\%)]  &   71(21\%)  & 66(19\%) & 6(1.8\%) &   72(21\%)\\
\hline
Totals & 20641 & 5105(25\%)[3388(16\%)] & 5685(27\%) & 5400(26\%) & 434(2.1\%)   & 5834(28\%) \\
\hline
\hline
\end{tabular}
\label{t1}
\end{table*}

\subsection{Comparison between the different AGN classification methods}
\label{compare}
The aim of this section is to compare the different methods used in this work to identify AGNs in order to find common objects between the different AGN selection criteria.

We start by comparing BPT and WHAN approachs. From the total of 5105 AGN+Composite galaxies identified in BPT we found 3307 are also WHAN AGNs. From the remaining 1798 BPT AGNs, 1717  objects are Retired/Passive objects according to WHAN and 81 have the WHAN star-formation category. These latter galaxies are very close ($<$0.2 dex) to the threshold log([NII]/H$\alpha$)=-0.4 adopted by WHAN to separate  star-forming galaxies from AGNs.

In the case of WHAN diagram there are 5685 AGNs, from these objects 2378 had not been classified as AGNs by BPT approach. From these galaxies 591 are above the \citet{Kauffmann2003} threshold but have a bad S/N relation. Also there are 1787 objects below this limit, with 1363 galaxies having S/N$>$3 in the four lines used for the BPT diagram, the distance of these galaxies to \citet{Kauffmann2003} demarcation line is lower than 0.2 dex. 

With regards to the WISE classification, we find that from 434 WISE AGNs, 366 are also Optical AGNs (BPT or WHAN). Therefore, there are only 68 WISE AGNs not classified as active objects in the optical from which 10 are Retired/passive galaxies, 52 are star-forming in either BPT or WHAN and the remaining 6 objects have low signal-to-noise relation in the optical lines.

\subsubsection{Retired/Passive galaxies in the BPT diagram}

 It has been shown that BPT diagram may not be able to distinguish between galaxies containing a weak AGN and Retired galaxies which have stopped forming stars and have emission associated to HOLMES \citep{Stasinska2008}. Therefore it is important to explore the Retired/Passive category in BPT classification.
For BPT AGNs there are 1717 galaxies classified as Retired/Passive objects according to WHAN diagram. \citet{Cluver2014} show that W2–W3 colour may be used to select galaxies with and without star formation, therefore it is interesting to study mid-Ir colour of BPT AGNs classified as Retired/Passive by WHAN. From these objects, 1542 (90\%) have W2-W3$<$2.5 consistent with the results of lineless galaxies suggested by \citet{Herpich2016}, and 728 (42\%) are below the most conservative limit W2-W3=1.5 adopted by \citet{Cluver2014}. 
Moreover if we study WISE AGNs in small galaxy systems we find that 10 galaxies are in the Retired/Passive area of the  WHAN diagram, this number correspond to only the 2\% of our mid-IR active sample. 
These results reinforce the category of Retired/passive galaxies as non active objects. For this reason in the forthcoming analysis we have removed 1717 Retired/passive galaxies from our sample of 5105 BPT AGNs. 
Therefore the final number of BPT AGNs is 3388, 3017 in pair systems, 328 in triplets and 43 active objects in groups.

Based on the previous analysis in what follows, we consider as AGNs, galaxies classified as active objects by at least one of the methods under study, i.e BPT without Retired/Pasive (hereafter BPT), WHAN or WISE. 
Under this combined criteria we have a total of 5834 AGNs in our sample, from these objects 5191 are in pairs, 571 in triples and 72 in groups. 

 A summary of the previous section are provided in table \ref{t1} where we show the total number of AGNs in pairs, triplets and groups, classified either by BPT, WHAN or WISE. We also include the number of Optical AGNs considering those AGNs classified by BPT or WHAN excluding WISE AGNs and the total number of AGNs selected with the combined criteria as described previously. 
 
 \begin{figure}
  \centering
  \includegraphics[width=.48\textwidth]{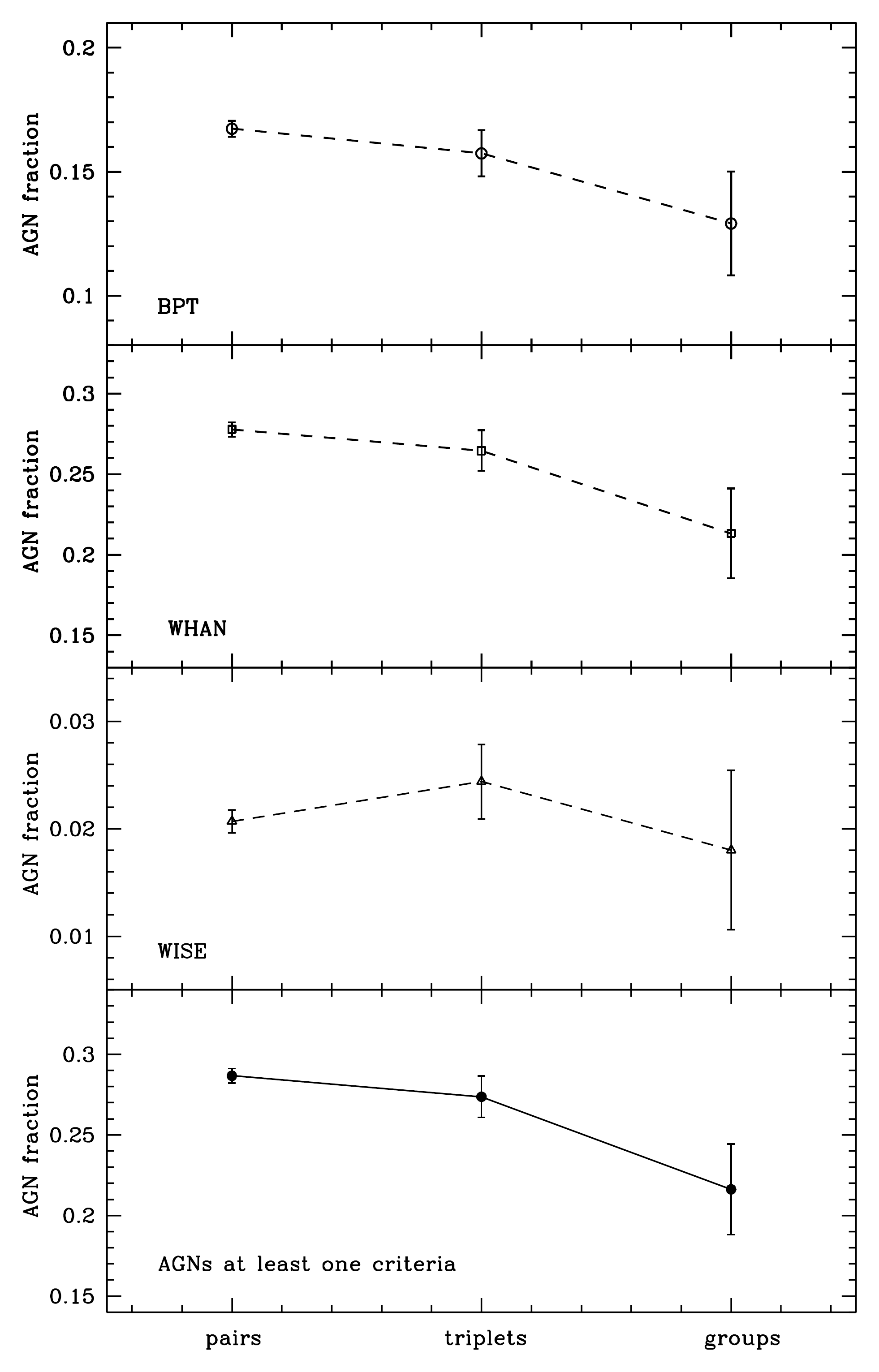}
  \caption{AGNs fraction for pairs triplets and groups. From top to bottom, AGNs selected according to BPT, WHAN  and WISE criteria. The bottom panel represent the fraction of AGNs selected by at least one of these methods. Error bars corresponds to Poisson errors.} 
\label{AGNfrac}
\end{figure}

\subsection{AGN fraction in small galaxy systems}
\label{frac}
 
 \begin{table*}
\centering
\caption{Number of galaxies in pairs, triplets and groups and number(fraction) of powerful AGNs in these systems according to BPT, WHAN or WISE AGN classification.  We also include the number of Optical powerful AGNs.} 
\begin{tabular}{lccccccc}
\hline\hline\noalign{\smallskip}
Name & powerful BPT  & regular BPT & powerful WHAN & regular WHAN  & powerful WISE &regular WISE \\
\hline\noalign{\smallskip}
Pairs &  2099(11\%) & 937(5.1\%) & 3559(20\%) & 1502(8.2\%) & 115(0.6\%)& 262(1.4\%)\\
Triplets &  224(11\%)  &  86(4.1\%) & 367(18\%) &  186(8.9\%) &  17(0.8\%)& 34(1.6\%)\\
Groups  & 23(6.9\%)  &   19(5.7\%)  & 41(12\%) & 30(9.0\%) &   2(0.6\%)& 4(1.2\%)\\
\hline
Totals &  2346(11\%) & 1042(5\%) & 3967(19\%) & 1718(8.3\%)   & 134(0.6\%)& 300(1.5\%) \\
\hline
\hline
\end{tabular}
\label{t2}
\end{table*}

In this section we consider the fraction of active galaxies selected with the different methods used in this work, as a function of the number of members in small galaxy systems. To this end  we use the information given in table \ref{t1} to plot Fig. \ref{AGNfrac}. From this figure it can be appreciated that for BPT selection, the AGN fraction is similar in pairs and triplets (17\% and 16\% respectively), and for groups there is a decreasing trend of the AGN fraction (which drops to 13\%). If we consider WHAN AGNs, we find similar trends with pairs and triplets having an AGN fraction of 28\% and 26\%, respectively, while for groups, the fraction is 21\%. In the selection of WISE AGNs in small systems we found a similar trend than for optical selected AGNs with pairs having an AGN fraction of 2\%, triplets 2.4\%  and groups 1.8\%. Nevertheless, as can be appreciated in Fig. \ref{AGNfrac} in all cases the error in the fraction of AGNs in groups is high given the small size of this sample. Fig. \ref{AGNfrac} bottom panel also shows the  tendency of the  AGNs fraction selected with our combined criteria. Again, pairs an triplets present a similar fraction of active objects (28\% and  27\%) and for groups the percentage is lower (21\%). These results suggest that the AGN fraction in pairs and triplets is always higher than the fraction of active objects in groups, regardless the adopted classification scheme.

As a complementary analysis, we also  identify powerful and regular AGNs and study the fraction of active objects in pairs, triplets and groups. For BPT selected AGNs,  we focus on the dust-corrected luminosity of the [OIII] line (L[OIII]) as a tracer of the AGN nuclear activity. This line was corrected for optical reddening using the Balmer decrement and the obscuration curve of \citet{Calzetti2000}.
The [OIII] line  is  one  of  the  strongest  narrow   emission  lines  in  optically obscured  AGNs  and  has  very  low  contamination  by  contributions  of  star  formation  in  the  host  galaxy \citep{Kauffmann2003}. We divide the samples into powerful and regular AGNs, considering the threshold Lum[OIII]=$10^7 L_\odot$. The same limit for strong AGNs was also defined by \citet{Kauffmann2003}. 
To select powerful and regular WHAN AGNs, we use the definition of strong and weak AGNs based on the equivalent width of H$\alpha$ line \citep{CidFernandes2011}. Therefore strong AGN has EW(H$\alpha$)$>$6 and for weak AGN the range is 3$<$\rm  EW(\rm H$\alpha$)$<$6. 
For WISE, we select as powerful AGNs those objects within \citet{Mateos2012} wedge, this region corresponds to luminous AGNs based on X-ray data. Objects outside this zone are considered as regular mid-IR AGNs.

\begin{figure}
  \centering
  \includegraphics[width=.48\textwidth]{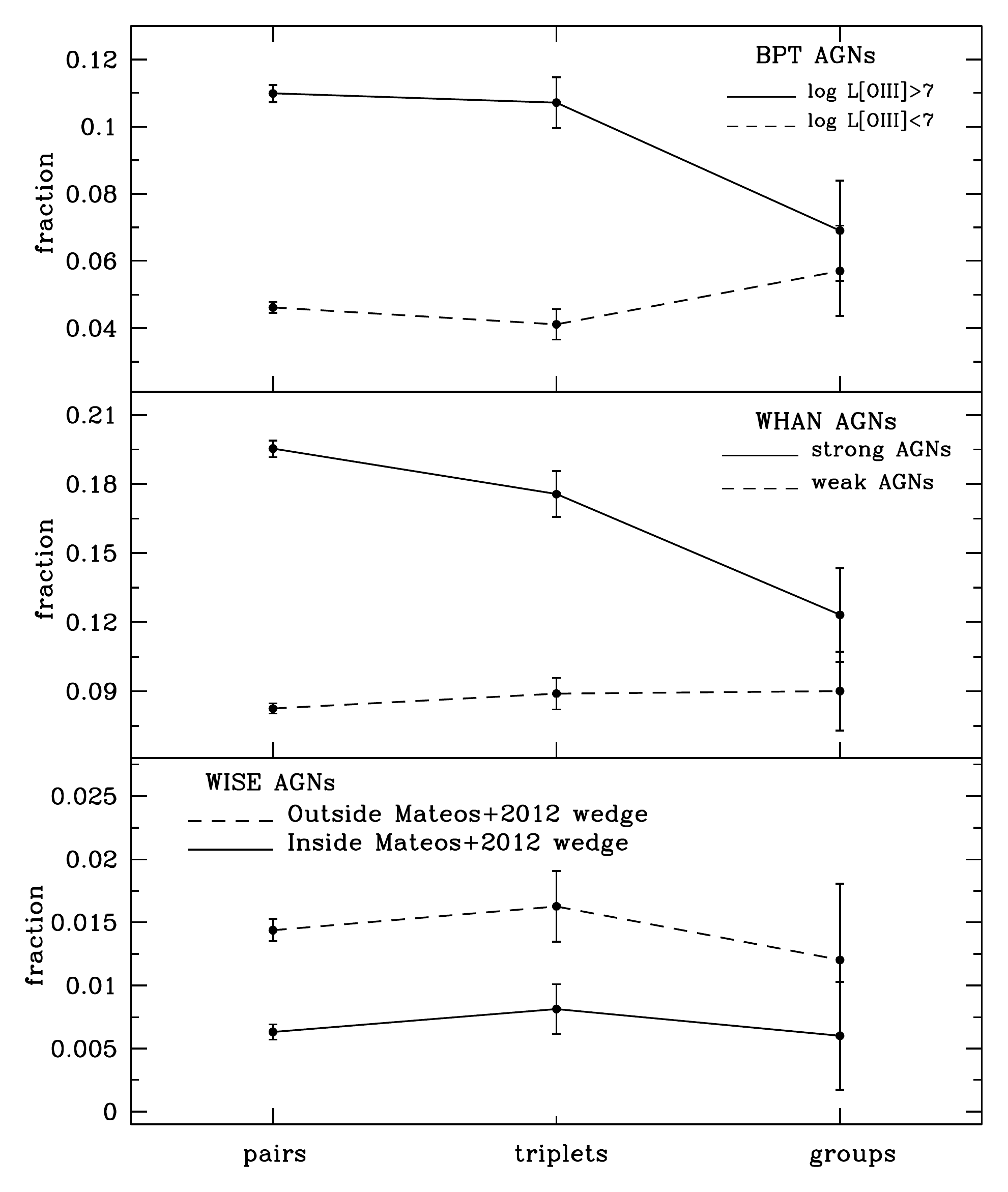}
  \caption{Fraction of powerful and regular AGNs for pairs, triplets and groups. From top to bottom BPT, WHAN and WISE AGNs. Error bars correspond to Poisson errors.} 
\label{AGNpowfrac}
\end{figure}

In table \ref{t2} we show the number and fraction of powerful and regular AGN selected according to the criteria described above. To better interpret these numbers in Fig. \ref{AGNpowfrac} we plot the fraction of powerful and regular AGNs in pairs, triplets and groups. It can be appreciated that for optically selected AGNs there is a higher fraction of powerful AGNs in pairs and triplets than in groups. Moreover the fraction of powerful AGNs in pair and triplets is about twice the fraction of regular AGNs but in groups the fractions of powerful and regular BPT/WHAN AGNs are similar. For WISE AGNs the fraction of regular AGNs is higher than the powerful AGNs despite the number of members in the system. Nevertheless for groups these fractions are similar within the errors given the small number of objects in these samples.

\subsection{Host properties}
\label{hosts}
In this section we study the fraction of AGNs in pair, triplets and groups depending on host properties.  To this end we use the \texttt{galSpec} galaxy properties from MPA-JHU emission line. From this catalogue we consider as a spectral indicator of the stellar population mean age the strength of the $4000$ \AA{} break ($\rm D_n(4000)$) defined as the ratio of the average flux densities in the narrow continuum bands 3850-3950 \AA{} and 4000-4100 \AA{} \citep{Balogh1999}. We also use the total stellar masses ($\rm M_*$) calculated from the photometry \citep{Kauffmann2003}.  We consider $(\rm M_g-\rm M_r)$ galaxy colour and the concentration index $\rm C=\rm r_{\rm 90}/\rm r_{\rm 50}$, where $\rm r_{\rm 90}$ and $\rm r_{\rm 50}$ are the radii containing 90\% and 50\% of the Petrosian galaxy light in the r band. This parameter is a suitable indicator of galaxy morphology: early type galaxies have $\rm C>2.6$ while late type galaxies have typically $\rm C<2.6$ \citep{Strateva2001}

\begin{figure}
  \centering
  \includegraphics[width=.45\textwidth]{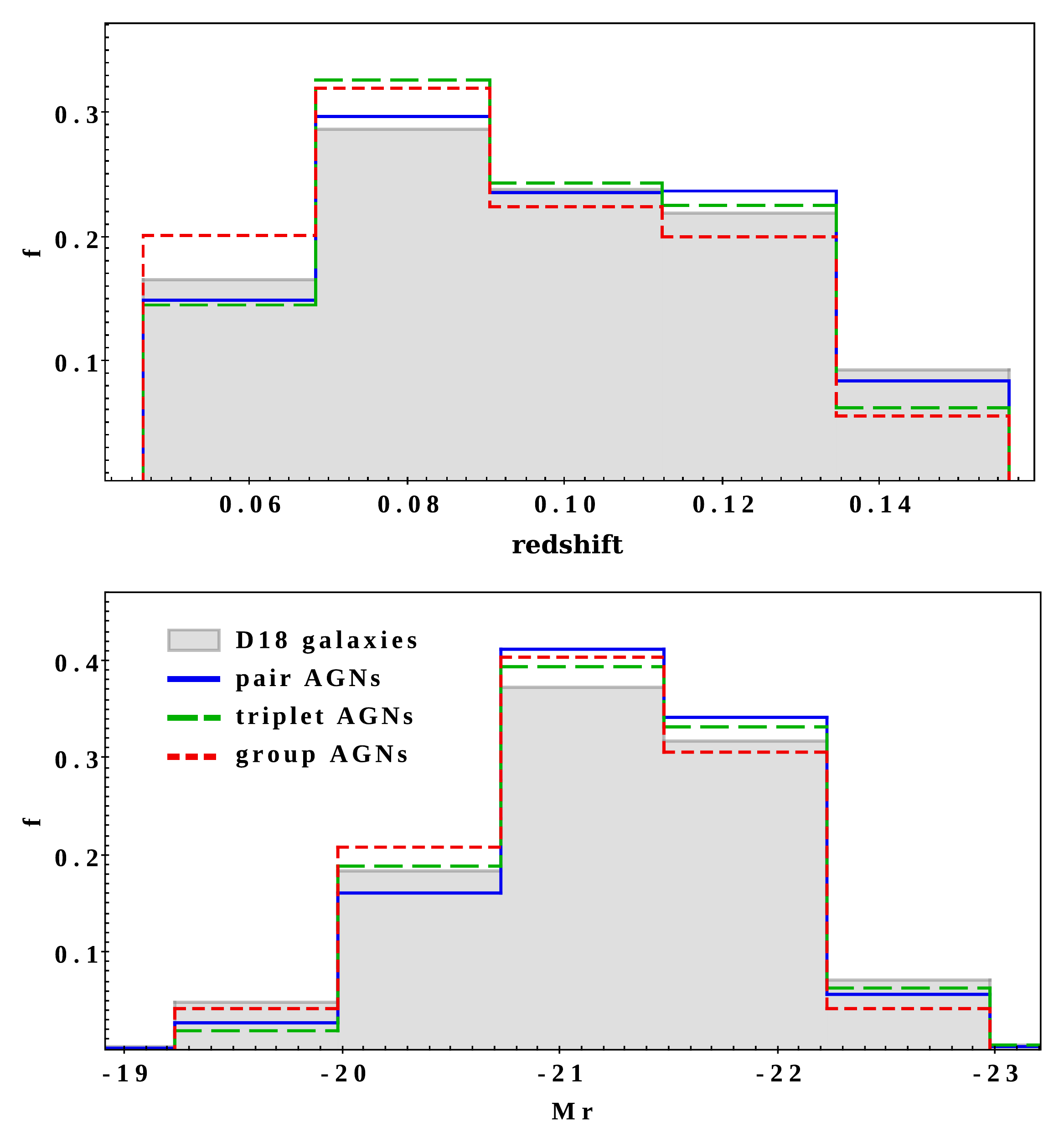}
  \caption{Normalised  distribution  of  redshift  (top)  and    r-band absolute magnitude Mr (bottom)  for  AGN host galaxies in pairs (solid), triplets (dashed) and groups (dotted). The shaded region corresponds to the distribution of all galaxies in D18 sample.} 
\label{z_Mr_dist_host}
\end{figure}

In Fig. \ref{z_Mr_dist_host} we show the redshift and absolute magnitude distributions of the AGN host galaxies in pairs, triplets and groups. Also we plot the distribution of all spectroscopic galaxies in D18 sample, it is worth to notice that there are no difference in the distribution of redshift nor magnitude for pairs, triplets and groups in D18 sample.  As can be seen, the samples show similar distributions  (p$>$0.05, for a Kolmogorov-Smirnov test), so that our results are not expected to be biased for differences in neither redshift nor magnitude of galaxies.

\begin{figure}
  \centering
  \includegraphics[width=.47\textwidth]{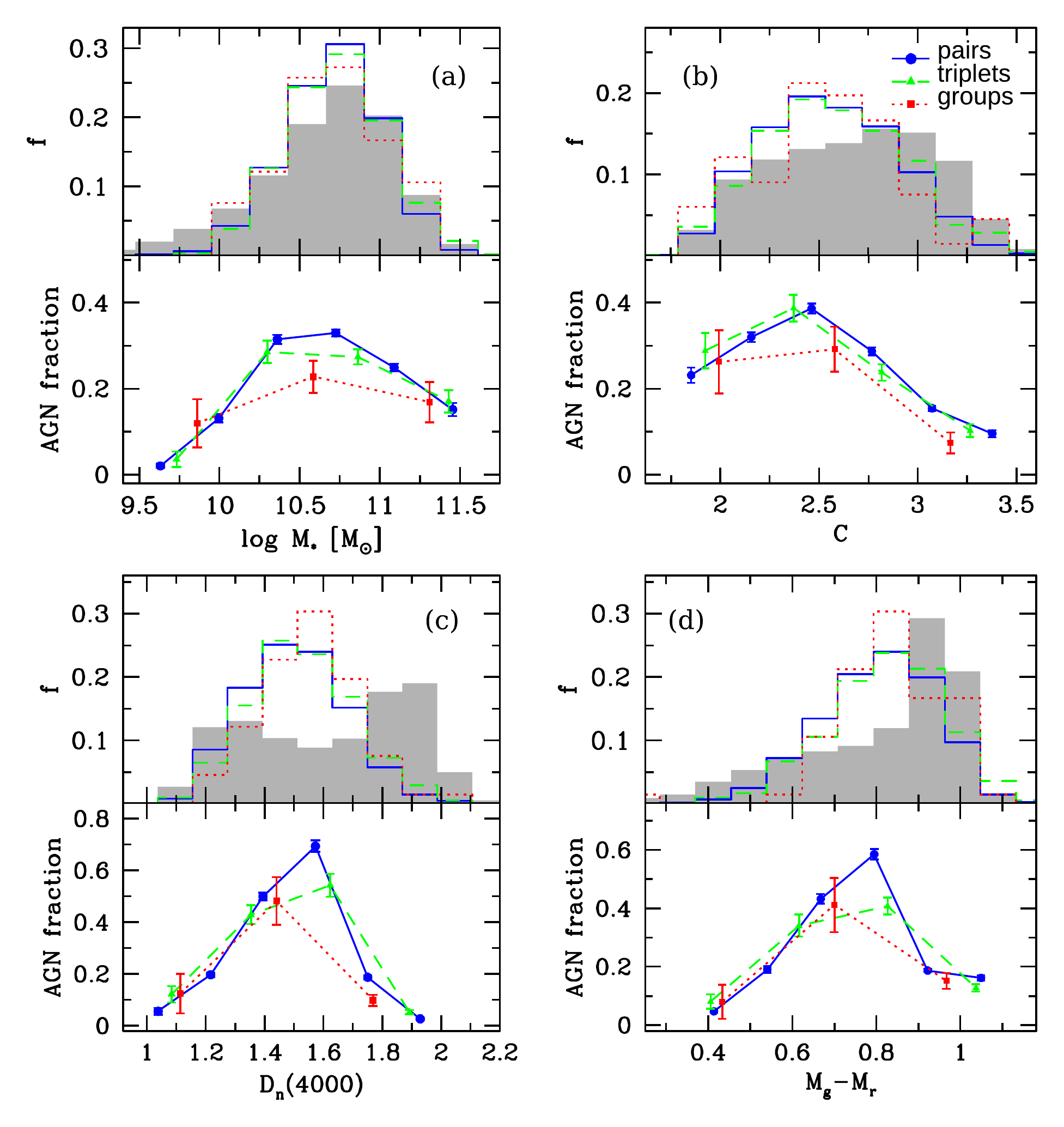}
  \caption{Normalised distribution of Optical AGN host properties (top panels) and AGN fraction as a function of different host properties (bottom panels), for pairs (solid), triplets (dashed) and groups (dotted). (a) Stellar mass $\rm M_*$, (b) concentration index C, (c) $\rm D_n(4000)$ index and (d) $\rm M_g-\rm M_r$  colour. The shaded histograms represent the distribution of the total sample. Error bars correspond to Poisson errors.} 
\label{AGN_frac_host}
\end{figure}

We define as Optical AGNs, those galaxies considered as AGNs by either BPT or WHAN classification in order to distinguish them from mid-IR WISE AGN selection. In
Fig. \ref{AGN_frac_host}  we plot the distribution of the stellar mass content $\rm M_*$, concentration index C, $\rm D_n(4000)$ index as stellar population age indicator and ($\rm M_g-\rm M_r$) colour, for Optical AGN host galaxies in pairs, triplets and groups. We also include the distributions for the total sample of spectroscopic galaxies in D18 small galaxy systems.  From these figures it can be appreciated that AGN hosts in groups have slightly redder colours and older stellar populations than the hosts in pairs and triplets. Note, however, that there is not a bimodal distribution as present for the global population of galaxies.  The $\rm M_*$ distribution of all the samples are rather similar and AGNs prefer less concentrated galaxies as can be appreciated from the C distributions. In Fig. \ref{AGN_frac_host} we also show the AGN fraction as a function of $\rm M_*$, C, $\rm D_n(4000)$ and  $(\rm M_g-\rm M_r)$, for pairs, triplets and groups. The fraction trends are similar despite the different number of system members, decreasing toward blue/red and old/young stellar population hosts. Despite the population of these type of hosts in pair, triplets and groups in D18, our results show that AGNs in small galaxy systems prefer hosts with intermediate stellar population properties.  

\begin{figure}
  \centering
  \includegraphics[width=.47\textwidth]{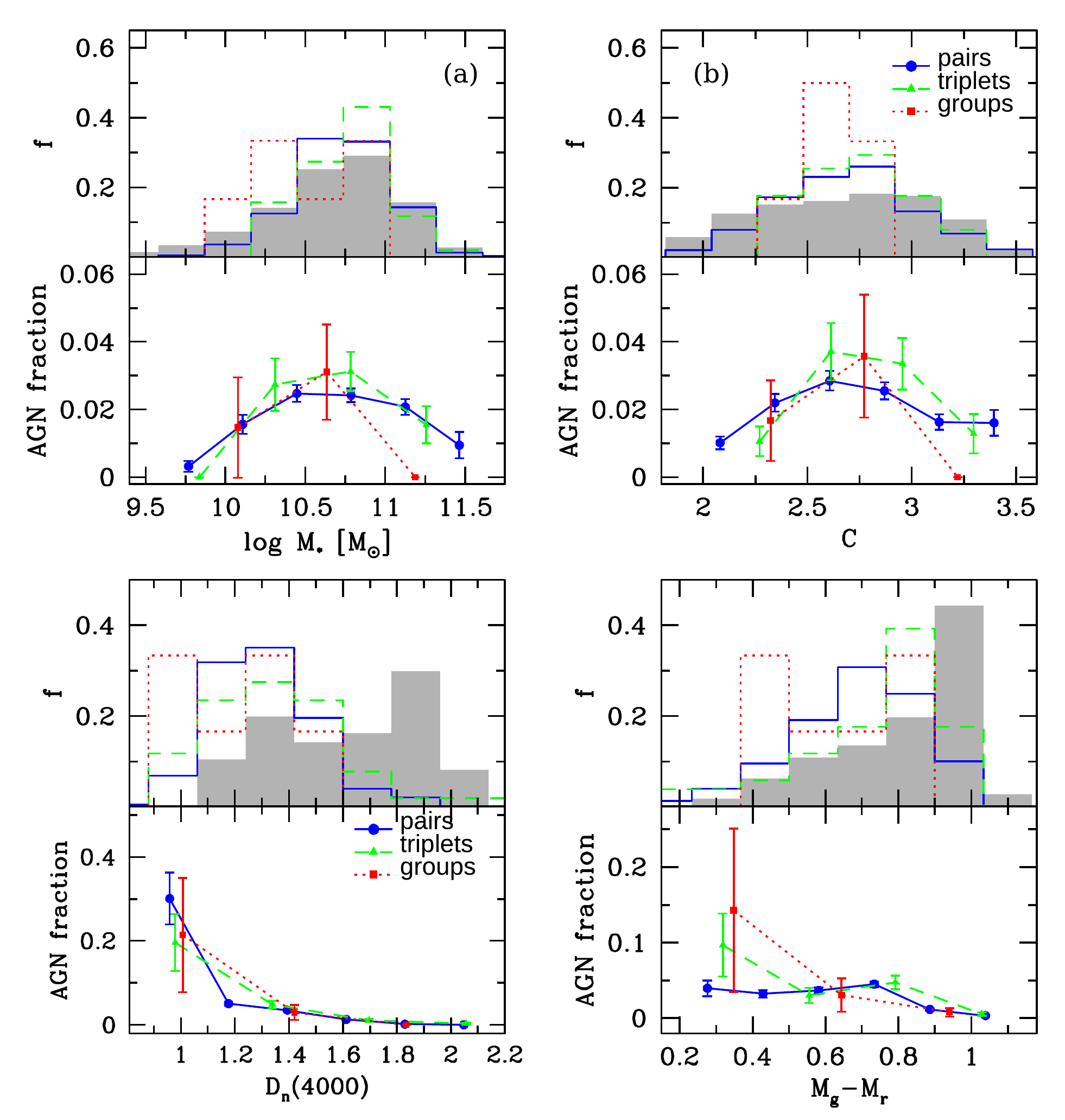}
  \caption{Normalised distribution of WISE AGN host properties (top panels) and AGN fraction as a function of different host properties (bottom panels), for pairs (solid), triplets (dashed) and groups (dotted). (a) Stellar mass $\rm M_*$, (b) concentration index C, (c) $\rm D_n(4000)$ index and (d) $\rm M_g-\rm M_r$  colour.The shaded histograms represent the distribution of the total sample. Error bars correspond to Poisson errors.}
\label{AGN_frac_host_WISE}
\end{figure}

We have also considered the properties of host galaxies of WISE AGNs and in Fig.\ref{AGN_frac_host_WISE} we show the main properties of these galaxies. We also include the distributions for the total sample of spectroscopic galaxies in D18 small galaxy systems. From this figure it can be seen the remarkable difference between the host properties  of mid-IR AGN and  Optical AGNs. The host galaxies of WISE AGNs in groups are less massive and concentrated, with young stellar populations and blue colours. For pairs and triplets the host properties are similar for Optical an WISE AGNs. Nevertheless the fraction of WISE AGNs increases toward bluer colours and younger stellar population hosts, despite the number of members in the system. 

\citet{Kauffmann2003} find that optical AGNs reside almost exclusively in massive galaxies with distributions of sizes, stellar surface mass densities and concentrations similar to those of ordinary early-type galaxies. We find similar results for our sample of Optical AGNs in small galaxy systems. In contrast, mid-IR AGNs in our sample follow a different trend compared to spectroscopic optical AGNs, preferentially hosted by galaxies with low stellar mass and young stellar populations. This effect is more evident for AGNs in groups. As described in section \ref{WHANAGN} the mid-IR AGN colour selection criteria identify objects where the AGN dominates over the host galaxy emission. Therefore, it should be expected a larger effect  in low mass galaxies, even if the \citet{Assef2018} selection criteria takes into account both colour and magnitude of the host. It is worth to notice that a similar result was found by \citet{Satyapal2014a} for WISE AGNs in bulgeless galaxies.

\subsection{Dependence on environment}
\label{environment}

 As shown by \citet{Duplancic2020}, the environment of small galaxy systems are diverse with groups residing in denser regions than pairs and triplets. Therefore,  we  explore the possibility that the different AGN properties of member galaxies in small systems are related  to the large scale environment hosting the system. In order to characterise the global environment we consider the surface density  parameter 
$$\Sigma_{\rm 5}={{\rm 5}\over{\pi\ \rm r_{\rm 5}^2}}$$ 
 
where $\rm r_{\rm 5}$ is the projected distance to the fifth neighbour galaxy  brighter than Mr$<$-20.5 with a radial velocity difference $\Delta \rm V< 1000 \kms$. This two-dimensional density estimator use the redshift information to reduce projection effects and is useful to characterise the local galaxy density. The advantage of this method with respect to a fixed distance based algorithm to count neighbours, is it adaptation to larger scales in lower-density regions which improves sensitivity and precision at low densities.

\begin{figure}
  \centering
  \includegraphics[width=.48\textwidth]{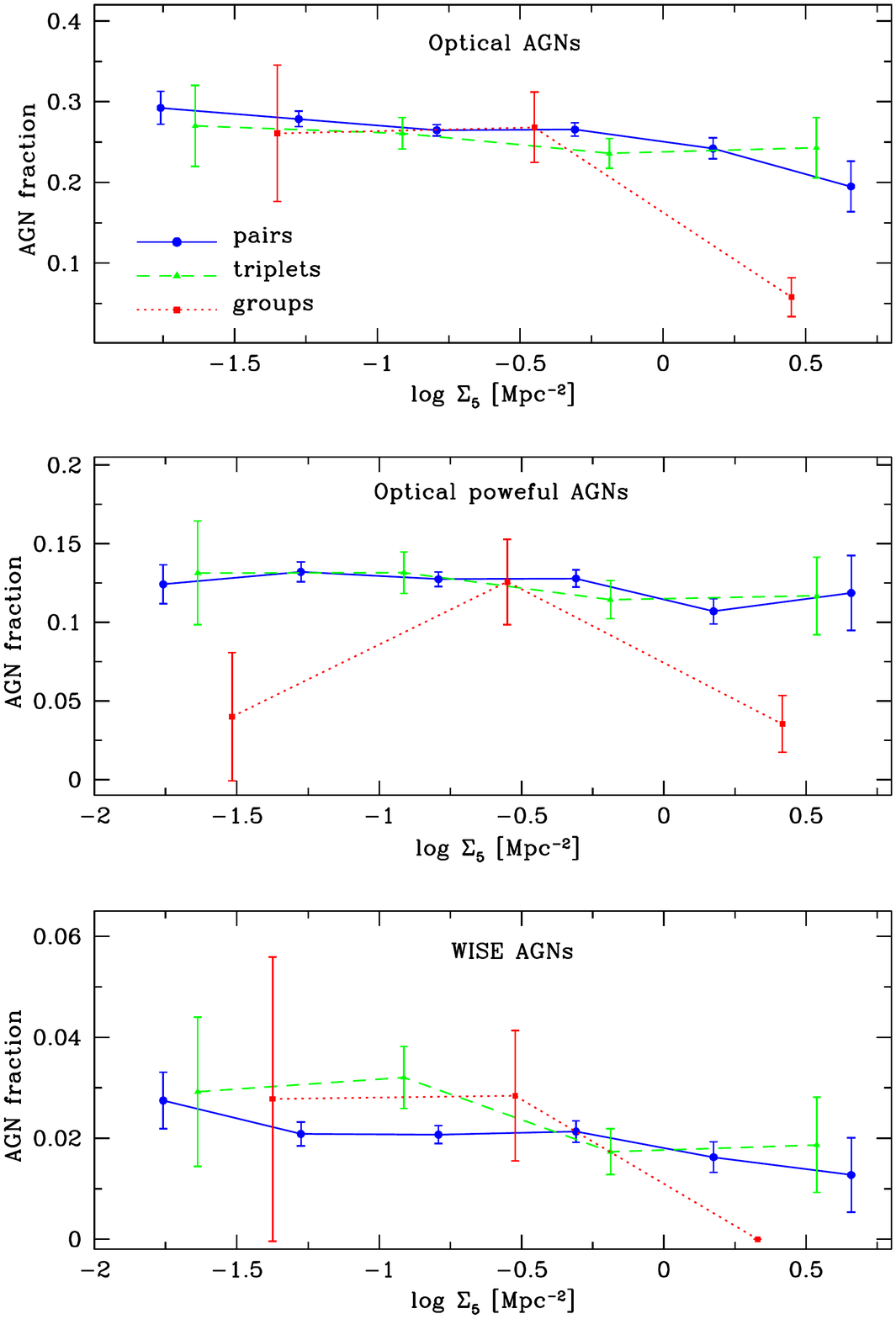}
  \caption{ AGN fraction as a function of $\Sigma_5$, for pairs (solid), triplets (dashed) and groups (dotted). From top to bottom, Optical AGNs, powerful AGNs and WISE AGNs. Error bars correspond to Poisson errors.} 
\label{AGNfrac_S5}
\end{figure}

In Fig. \ref{AGNfrac_S5} we explore the AGN fraction as a function of the density parameter $\Sigma_{\rm 5}$ for pairs, triplets and groups, and where Optical and WISE AGNs members are considered separately. Also we consider Optical powerful AGNs by requiring  either Lum[OIII]$>10^7 L_\odot$, or strong AGN class according to WHAN classification. 
From this figure it can be appreciated that for Optical AGNs there is a tendency for pairs to host a lower AGN fraction in globally higher density regions. On the other hand, the AGN fraction in triplets is constant for the entire density range while for groups there is a clear signal of a decreasing AGN fraction in systems residing in high density environments. For WISE AGNs in pairs triplets and groups we found similar trends (within the errors) than Optical AGNs. Nevertheless, WISE AGNs in groups tend to avoid higher density regions, with a null fraction of AGNs for the densest $\Sigma_5$ bin values, even for a considerable number of D18 group member galaxies residing in these regions.  It is worth to notice that D18 group galaxies are redder/older that galaxies in pairs and triplets, therefore the previous results are in agreement with \citet{Coldwell2009} who found that red AGN hosts inhabit environments less dense compared to non-active red galaxies.

For Optical powerful AGNs, pairs and triplets present a constant trend of the AGN fraction as a function of $\Sigma_5$. For groups, the distribution from higher to lower density regions shows an increment of the AGN fraction that drops rapidly for the lowest values of $\Sigma_5$. Therefore powerful AGNs in groups seems to require a more restricted density environment.

\begin{figure}
  \centering
  \includegraphics[width=.48\textwidth]{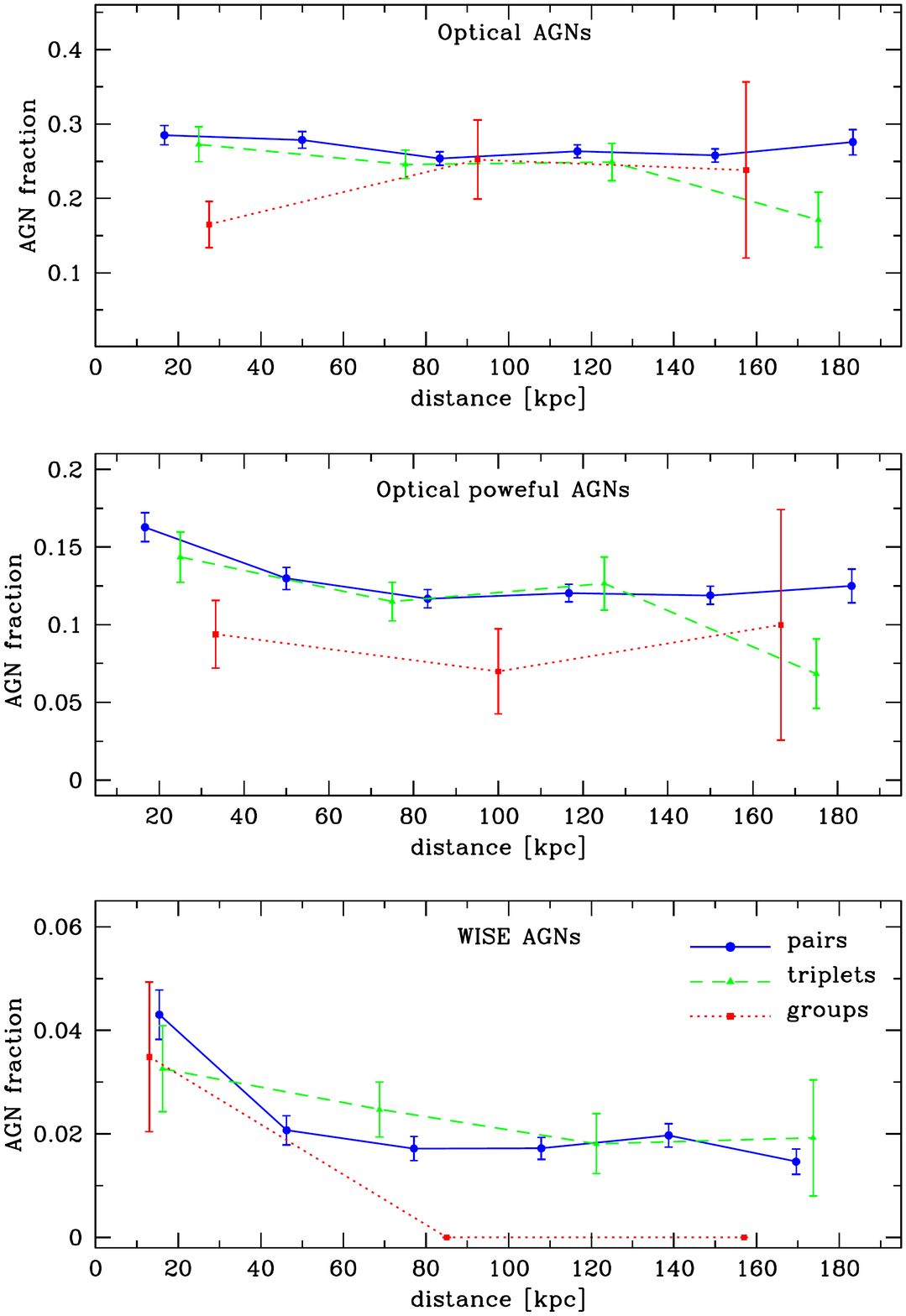}
   \caption{AGN fraction as a function of the distance to the nearest companion within the system for pairs (solid), triplets (dashed) and groups (dotted). From top to bottom, Optical AGNs, powerful AGNs and WISE AGNs. Error bars correspond to Poisson errors.} 
\label{AGNfrac_rpv}
\end{figure}

As a complementary analysis, we have studied the influence of the more local environment. To this end, we calculate the AGN fraction as a function of the distance to the nearest companion galaxy within the system, the results are shown in Fig. \ref{AGNfrac_rpv}. We study the distributions of this parameter for Optical and mid-IR WISE AGNs, and we also consider separately, Optically powerful AGNs as described previously. From this figure, it can be seen a relative constant trend for the Optical AGN fraction in pairs while triplets show a larger AGN fraction for galaxies with closer nearest neighbours.  On the other hand, it can be seen the decline of the AGN fraction for groups when considering separations lower than  100 kpc. Moreover powerful AGNs in groups present a constant trend, always showing a lower AGN fraction than pairs and triplets. 

 For WISE AGNs it can be appreciated a significant tendency of higher AGN fraction with decreasing distance to the nearest companion. Moreover, all WISE AGNs detected in groups have a companion closer than 20 kpc and with a similar  AGN fraction than triplets and groups,  despite the low number statistics. We find no WISE AGNs in groups with nearest companions at larger separations, even though either pairs, triplets or groups span a similar range of distance to the nearest neighbour. 
 Given the small number of group WISE AGNs sample (only 6 galaxies) we have estimated the probability that 6 galaxies selected at random from the sample of Optical AGNs have distances to the nearest neighbour closer than 20 kpc. By performing 100 realisations, we find this probability to be only 1\%, which provides confidence in our results.

\section{Discussion}
\label{disc}

 It can be argued that in systems with more than three main members, gas depletion caused by repeated interactions may drive most of the gas into the intragroup  medium  of  the system \citep{Verdes-Montenegro2001}. Also, transient phenomena such as shocks can prevent interstellar medium of galaxies to relax and flow convergently to the central regions of galaxies.  With no gas collimated onto the centre, nuclear activity is not likely to be triggered. These scenario may be influencing the AGN fraction in pairs, triplets and groups. For paired systems \citet{Alonso2007} identified BPT AGN hosts in a catalogue of  close pairs comprising systems with relative projected separations, rp $<$ 25 kpc h$^{-1}$ and relative radial velocities, $\Delta$V$<$ 350 km $s^{-1}$ within redshifts z$<$ 0.1. They found that 30\% of the galaxy pairs have one member exhibiting AGN activity. They also detected that 7\% of the close pairs have both members classified as AGNs. In a similar direction, \citet{Lambas2012} detected AGNs in close pair galaxies (i.e. pairs with rp$<$25 h$^{-1}$kpc  and $\Delta$V$<$350 km s$^{-1}$) from SDSS-DR7, finding that the 23.73\% are AGN-galaxy pairs and the  6.63 \% are AGN-AGN pairs. These AGN fractions are in agreement with the results of the present work where we find a fraction of 26\% of optically selected AGNs in our pair sample. Nevertheless \citet{Koulouridis2013} analyse the optical spectroscopy and X-ray imaging of neighbouring galaxies around a samples of Seyfert 1 and Seyfert 2 and find that the fraction of AGNs can be higher than 70\%, indicating a link between close interactions, and star formation/nuclear activity.
For galaxy triplets \citet{CostaDuarte2016} study a sample of 240 triplet galaxies finding a fraction of 23\% of BPT AGNs that are also strong/weak AGNs according to WHAN. This fraction is similar to the 25\% of AGNs in our triplet sample.   
For compact groups different works find AGN fractions ranging from 40\% to 60\% \citep{Coziol1998,Martinez2010,DeRosa2015}. Nevertheless \citet{Sohn2013} used emission-line ratio diagrams to identify AGN host galaxies finding that the AGN fraction of compact groups of galaxies ranges from 17\% to 42\% depending on the AGN classification method. They also found that the AGN fraction is not the highest among different galaxy environments. In the present work we find a fraction of only 20\% of AGNs in our galaxy groups.
It is worth to notice than \citet{Martinez2010} and \citet{Sohn2013} found no mid-infrared  AGN hosts in their sample while in the present work we found a low fraction (2\%) of WISE AGNs in groups. These AGNs are associated to low mass, blue, young stellar population galaxies with very close companions, residing  in  intermediate global density environment.  These results may be revealing that mid-IR AGNs in small galaxy systems are associated to a different population than spectroscopic Optical AGNs.

Our results can be interpreted following the studies by \citet{Popesso2006} who found a decreasing fraction of AGN for larger cluster velocity dispersion and number of members. This result can be related to the close galaxy interaction rate given its inverse cubic dependence on the system velocity dispersion. 
Following this line, several works report that different types of AGN avoid high density regions such as galaxy system centres \citep{Coldwell2006,Coldwell2014,Coldwell2017}. Instead, the average AGN environment is characterised by the presence of blue, disk-type and star forming galaxies which are frequent in the periphery of clusters and groups. Interestingly, AGN have also been found preferentially between merging cluster \citep{Soetching2002,Soetching2004} where the merger rate may also be higher if progenitor relative velocity is smaller.

The results found for pairs and triplets in our sample suggest that interactions can provide an efficient mechanism for feeding the central regions of galaxies, activating the AGN.
In this line, \citet{Alonso2007} performed a statistical analysis of AGN activity in galaxy pairs considering a very well defined sample of isolated AGN. The results show that AGN OIII luminosity is enhanced for hosts with strong interaction features (i.e. merging pairs) in comparison to a control sample of isolated AGNs. In addition, estimations of the mean accretion rates onto the black holes of AGNs in pairs compared to those in isolation show AGNs in close interacting pairs having more active black holes at a given host r-band luminosity or stellar mass content. Moreover, all AGNs in interactions present signs of active star formation indicating the  possibility that isolated AGNs with active black holes have experienced a recent merger.

\section{Summary and Conclusions}
\label{conc}

In this work we study AGNs in the sample of D18 galaxies in small systems. To this end we consider two methods to select optical AGNs, BPT and WHAN. Also we identify mid-IR AGNs by using WISE data. The main results of this study can be summarised in the following items.

\begin{itemize}
    \item  Regardless the adopted classification scheme, the AGN fraction in pairs and triplets is always higher than the fraction of active objects in groups.
    
    \item  For  optical  AGNs  there is a higher fraction of powerful AGNs in pairs and triplets than in groups. Moreover the fraction of powerful AGNs in pair and triplets is about twice the fraction of regular AGNs. On the other hand for groups the fractions of powerful and regular Optical AGNs are similar. For WISE AGNs the fraction of regular AGNs is higher than the powerful AGNs despite the number of members in the system. Nevertheless for triplets and groups these fractions are similar within the errors given the small number of objects in these samples.
    
    \item We study the host properties finding that for pairs and triplets the host of Optical (BPT/WHAN) an WISE AGNs are rater similar. For groups there is a remarkable difference between Optical and mid-IR AGNs, being the host  galaxies of WISE AGNs in groups less massive and concentrated, with young stellar populations and blue colours. However the fraction of WISE AGNs increases toward bluer colours and younger stellar population hosts, despite the number of members in the system.
    
    \item We consider the fraction of AGNs as a function of the density parameter $\Sigma_{\rm 5}$ for pairs, triplets and groups. Pairs present a tendency to host a lower AGN fraction in globally higher density regions. On the other hand, the AGN fraction in triplets is constant for the entire density range while for groups there is a clear signal of a decreasing AGN fraction in systems residing in high density environments. Moreover powerful AGNs in groups seems to require a more restricted density environment.
    
    \item  As a complementary analysis, we study the influence of a more local environment. To this end we calculate the AGN fraction as a function of the distance to the nearest companion within the system finding a constant trend for the AGN fraction in pairs while triplets show a larger AGN fraction for galaxies with close neighbours.  For groups there is a decline in the AGN fraction for galaxies with companions closer than 100 kpc. Regarding WISE AGNs, it can be appreciated an increasing trend of the AGN fraction for galaxies with a close companion. Moreover for galaxy groups all WISE AGNs have a companion closer than 20 kpc with a similar fraction despite the number of members in the system.
    
\end{itemize}

In D18 we study the variation of star forming indicators respect to the distance to the  nearest companion within the system finding an enhancement of star forming galaxies with companions closer than 100 kpc despite the number of members in the system. This may be an indication of an interaction-induced star formation activity due to recent close encounters.  Nevertheless pairs are more strongly star forming and bluer than triplet and group members. 
In \citet{Duplancic2020} we found that the differences in the properties of member galaxies in small systems are not only related to the existence of an extra galaxy  but also to the large scale environment inhabited by the system. 
 In the present work we find a decreasing fraction of AGNs with increasing number of members in small galaxy systems. We suggest repeated interactions  may trigger transient phenomena as shocks and  activate  the star formation suppressing  mechanisms more efficiently in galaxy groups than in triplets and pairs. Moreover we find a connection between the AGN fraction and global/local environmental density estimators.  In sum, the results of this work highlight the important role of interactions, besides the  global environment dependence, in the activation of the AGN phenomenon in small galaxy systems.
We also acknowledge the relevance of another important issue related to this work which is the AGN-starburst connection. This important topic will be analysed in a forthcoming paper.

\section*{Acknowledgments}
We thank the Referee for her/his very useful corrections comments and advise which helped to improve this paper.
This work was supported in part by the Consejo Nacional de Investigaciones Cient\'ificas y T\'ecnicas de la Rep\'ublica Argentina (CONICET) and Secretar\'ia de Ciencia y T\'ecnica de la Universidad Nacional de San Juan. Funding for the Sloan Digital Sky Survey IV has been provided by the Alfred P. Sloan Foundation, the U.S. Department of Energy Office of Science, and the Participating Institutions. SDSS-IV acknowledges support and resources from the Center for High-Performance Computing atthe University of Utah. The SDSS web site is www.sdss.org. SDSS-IV is managed by the Astrophysical Research Consortium for the Participating Institutions of the SDSS Collaboration including the Brazilian Participation Group, the Carnegie Institution for Science, Carnegie Mellon University, the Chilean Participation Group, the French Participation Group, Harvard-Smithsonian Center for Astrophysics, Instituto de Astrof\'isica de Canarias, The Johns Hopkins University, Kavli Institute for the Physics and Mathematics of the Universe (IPMU) / University of Tokyo, Lawrence Berkeley National Laboratory, Leibniz Institut f\"ur Astrophysik Potsdam (AIP),  Max-Planck-Institut f\"ur Astronomie (MPIA Heidelberg), Max-Planck-Institut f\"ur Astrophysik (MPA Garching), Max-Planck-Institut f\"ur Extraterrestrische Physik (MPE), National Astronomical Observatories of China, New Mexico State University, New York University, University of Notre Dame, Observat\'ario Nacional / MCTI, The Ohio State University, Pennsylvania State University, Shanghai Astronomical Observatory, United Kingdom Participation Group, Universidad Nacional Aut\'onoma de M\'exico, University of Arizona, University of Colorado Boulder, University of Oxford, University of Portsmouth, University of Utah, University of Virginia, University of Washington, University of Wisconsin, Vanderbilt University, and Yale University. This publication makes use of data products from the Wide-field Infrared Survey Explorer, which is a joint project of the University of California, Los Angeles, and the Jet Propulsion Laboratory/California Institute of Technology, funded by the National Aeronautics and Space Administration.

\section*{DATA AVAILABILITY}
The data underlying this article will be shared on reasonable request to the corresponding author.
\bibliographystyle{mnras.bst}
\bibliography{Bib}{}

\end{document}